\documentclass{mn2e}
\usepackage{psfig}
\usepackage[authoryear]{natbib}

\newcommand\xmm{{\it XMM-Newton}}

\newcommand\balqsoa{XWAS~J005007.3-521506} %X21211_017
\newcommand\balqsob{XWAS~J005044.2-515950} %X21211_104
\newcommand\balqsoc{XWAS~J015239.2-133703} %X00955_00149
\newcommand\balqsod{XWAS~J015257.2-140458} %X01135_00026
\newcommand\balqsoe{XWAS~J022645.4-043615} %X21565_00025
\newcommand\balqsof{XWAS~J050754.2-372726} %X00912_00137
\newcommand\balqsoashort{XWAS~J005007} %X21211_017
\newcommand\balqsobshort{XWAS~J005044} %X21211_104
\newcommand\balqsocshort{XWAS~J015239} %X00955_00149
\newcommand\balqsodshort{XWAS~J015257} %X01135_00026
\newcommand\balqsoeshort{XWAS~J022645} %X21565_00025
\newcommand\balqsofshort{XWAS~J050754} %X00912_00137

\title{X-ray selected BALQSOs}
\author[Page, et al.]
{M.J. Page\(^{1}\), F.J. Carrera\(^{2}\), M. Ceballos\(^{2}\), 
A. Corral\(^{3}\), J. Ebrero\(^{4}\), P. Esquej\(^{5}\), \and
M. Krumpe\(^{6}\), S. Mateos\(^{2}\), S. Rosen\(^{7}\), 
A. Schwope\(^{6}\), A. Streblyanska\(^{8}\), \and 
M. Symeonidis\(^{1}\), J.A. Tedds\(^{7}\), M.G. Watson\(^{7}\)\\
\\
\(^{1}\)Mullard Space Science Laboratory, University College London,
Holmbury St Mary, Dorking, Surrey RH5 6NT, UK.\\
\(^{2}\)Instituto de F\'\i sica de Cantabria 
(CSIC--Universidad de Cantabria), 39005
Santander, Spain.\\
\(^{3}\)IAASARS, National Observatory of Athens, 15236 Penteli, Greece\\
\(^{4}\)XMM-Newton Science Operations Centre, ESA, Villafranca del Castillo, Apartado 78, 28691 Villanueva de la Ca\~nada, Spain\\
\(^{5}\)Herschel Science Centre, ESA, Villafranca del Castillo, Apartado 78, 28691 Villanueva de la Ca\~nada, Spain\\
\(^{6}\)Leibniz-Institut f\"ur Astrophysik Potsdam (AIP), An der Sternwarte 16, 14482 Potsdam, Germany\\
\(^{7}\)Department of Physics and Astronomy, University of Leicester, Leicester, LE1 7RH, UK\\
\(^{8}\)Instituto de Astrof\/isica de Canarias (IAC), 38200 La Laguna, Tenerife, Spain\\
}

\date{}

\begin{document}
\maketitle

\begin{abstract}
  We study a sample of six X-ray selected broad absorption line (BAL)
  quasi-stellar objects (QSOs) from the {\em XMM-Newton} Wide Angle
  Survey. All six objects are classified as BALQSOs using the classic
  balnicity index, and together they form the largest sample of X-ray
  selected BALQSOs. We find evidence for absorption in the X-ray
  spectra of all six objects. 
  An ionized absorption model applied
  to an X-ray spectral shape that would be typical for non-BAL QSOs 
(a power law with energy index $\alpha=0.98$)  provides acceptable fits to
  the X-ray spectra of all six objects. 
The optical to X-ray spectral indices, $\alpha_{OX}$, of the X-ray
  selected BALQSOs, have a mean value of 
$\langle \alpha_{OX} \rangle = 1.69\pm 0.05$, which is similar to that 
found for X-ray selected and optically selected non-BAL QSOs of similar 
ultraviolet luminosity. In contrast, optically-selected BALQSOs typically 
have much larger $\alpha_{OX}$ and so are characterised as being X-ray weak.
The results imply that X-ray selection yields intrinsically X-ray bright
  BALQSOs, but their X-ray spectra are absorbed by a similar degree to that 
seen in optically-selected BALQSO samples; X-ray absorption
  appears to be ubiquitous in BALQSOs, but X-ray weakness is not. 
  We argue that BALQSOs sit at one 
  end of a spectrum of X-ray absorption properties in QSOs related to the
  degree of ultraviolet absorption in C\,IV\,1550\AA.
\end{abstract}

\section{Introduction}

Broad absorption line (BAL) quasi-stellar objects (QSOs) are a subset
of QSOs which show blue-shifted absorption lines with widths
$>2000$~km~s$^{-1}$ in their rest-frame ultraviolet spectra. Most are
identified by absorption due to C\,IV, observed to the blue of the
C\,IV 1550~\AA\, broad emission line. BALQSOs represent between 5 and
20 per~cent of the QSO population \citep{trump06, gibson09, scaringi09}, 
with the fraction somewhat
dependent on the criteria used to define BALQSOs.  It
was realised during the 1990s that BALQSOs were systematically fainter
in the X-ray band than QSOs without BALs \citep{green95,green96}.
Since the launches of {\em XMM-Newton} and {\em Chandra}, considerable
advances have been made in characterising the X-ray properties of
optically-selected
BALQSOs \citep[e.g. ][]{gallagher06, morabito14}, with strong X-ray 
absorption appearing to be an ubiquitous characteristic of
optically-selected 
BALQSOs.

As might be expected, given their X-ray faintness, BALQSOs are rarely
discovered through X-ray sky surveys. Exceptions include one BALQSO
found in the Chandra Deep Field South \citep{giacconi01}, and two
discovered during the AXIS survey \citep{barcons02}.  The largest
sample of X-ray selected BALQSOs comes from the 1$^{H}$~Deep Field,
which yielded four such objects \citep{blustin08}.  The properties of the
four 1$^{H}$ BALQSOs and the CDFS BALQSO were studied in detail by
\citet{blustin08}. They found that in terms of their 
rest-frame 2500~\AA\ to 2~keV spectral slopes ($\alpha_{ox}$) 
the X-ray-selected BALQSOs were a
little less X-ray faint than optically-selected BALQSO samples, 
but
still faint in the X-ray compared to X-ray selected non-BAL QSO samples. 
From the
X-ray spectral shapes, \citet{blustin08} inferred that the
X-ray-selected BALQSOs had similar levels of X-ray absorption to those
seen in optically-selected samples of BALQSOs.

Somewhat contradictory results were presented by \citet{giustini08},
who 
obtained a sample 
of 54 X-ray detected 
BALQSOs by cross-correlating the
Sloan Digital Sky Survey (SDSS) Data Release 5 (DR5) QSO catalogue
\citep{schneider07} with the Second {\em XMM-Newton} Serendipitous
X-ray Source Catalogue \citep[2XMM; ][]{watson09}. \citet{giustini08}
found that one third of the BALQSOs in their sample showed little or
no X-ray absorption in their X-ray spectra, and that the distribution
of $\alpha_{OX}$ of their sample was indistinguishable from that of
non-BAL
QSOs, implying that the BALQSOs in their sample are not X-ray
weak.

Subsequently, \citet{streblyanska10} compiled a sample of 88
X-ray-detected BALQSOs by cross-correlating the 2XMM catalogue with
BALQSOs in the NASA Extragalactic Database (NED). They found that when
the X-ray absorption is modelled as ionized gas, as opposed to the
neutral absorber model considered by \citet{giustini08}, 90 per cent
of the X-ray detected BALQSOs showed X-ray
absorption. 
\citet{streblyanska10} did not examine the $\alpha_{OX}$
distribution of their BALQSO sample.

It is notable that the X-ray properties found for samples of 
BALQSOs may depend on the criterion that was used to 
identify the BALQSOs from their
rest-frame UV spectra.
A quantitative criterion for classifying an object as a
BALQSO was introduced in the form of the balnicity index (BI) by
\citet{weymann91}.  For a full definition of balnicity index, see
Appendix A of \citet{weymann91}, but in simple terms it is
approximately the equivalent width in km~s$^{-1}$ of continuous
troughs of absorption, with outflow velocities between 3000 and 30000
km~s$^{-1}$ to the blue of C\,IV $\lambda$1550, and from which the 
first 2000 km~s$^{-1}$ velocity interval of each
absorption trough is excluded from the calculation. 
Thus BI$>0$ implies at least one absorption trough with a 
width of $>2000$~km~s$^{-1}$. 
According to \citet{weymann91}, BI$>0$ is required
to classify an object as a BALQSO. 
An alternative criterion, known as
the absorption index (AI), was introduced by \citet{trump06}, 
based on an earlier 
proposal by \citet{hall02}. In
essence, the AI is also an equivalent width measurement, but includes
the contribution of absorption troughs with outflow velocities below
3000~km~s$^{-1}$, and requires a minimum absorption trough width of
only 1000~km~s$^{-1}$ to yield a positive AI. Thus BI$>0$ is a more
conservative criterion for the classification of BALQSOs than AI$>0$.
Prior to 2006, BALQSOs studied in the X-ray were usually selected by
BI, as was the sample studied by \citet{blustin08}. However the BALQSO
sample of \citet{giustini08} was based on AI. \citet{streblyanska10}
selected the sample on AI, but also presented BI values for their
sample. They found
that while the sources in their sample with BI$>0$ were almost always
absorbed, the sources in their sample with BI=0 typically did not have
significant X-ray absorption. 

In this paper we present a new sample of X-ray selected BALQSOs, which
is larger and brighter in X-rays than the sample presented by
\citet{blustin08}. We use the new sample to investigate the X-ray
properties of such objects, in particular their optical to X-ray
spectral indices and the incidence of X-ray absorbing gas.
Throughout, we use the phrase ``X-ray selected BALQSOs''
specifically for objects which were discovered first as X-ray sources,
and then found to be BALQSOs in the subsequent optical spectroscopic
follow up. 
We define power law spectral indices $\alpha$ such that $f_{\nu}\propto
\nu^{-\alpha}$.
We have assumed cosmological parameters
$H_{0}=70$~km~s$^{-1}$~Mpc$^{-1}$, $\Omega_{\Lambda}=0.7$ and
$\Omega_{\rm m}=0.3$. Unless stated otherwise, all uncertainties are
given at 1\,$\sigma$.

\begin{figure}
\begin{center}
\leavevmode
\hspace{-8mm}
\psfig{figure=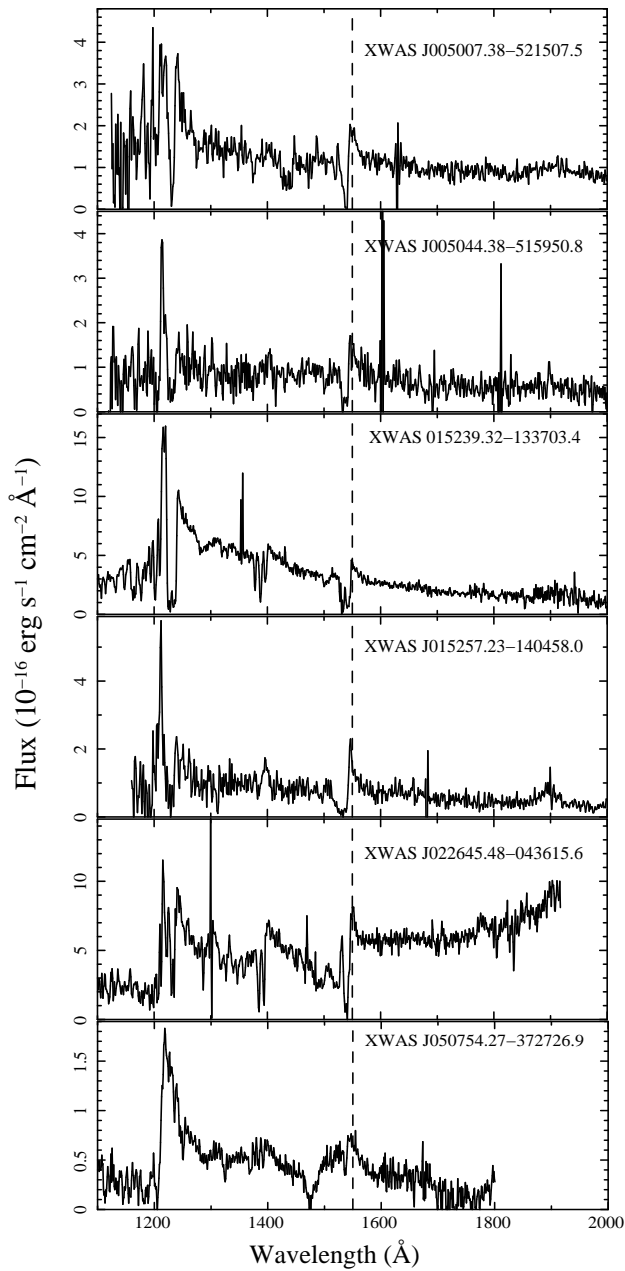,width=88mm}
\caption{Rest-frame ultraviolet spectra of the X-ray selected BALQSOs. 
The dashed line marks the rest-frame wavelength of C\,IV, 1550\AA.}
\label{fig:grest}
\end{center}
\end{figure}

\section{Sample and UV absorption properties}

\begin{table*}
\begin{center}
\caption{List of XWAS BALQSOs and their ultraviolet absorption properties.}
\label{tab:opticalproperties}
\begin{tabular}{cccccc}
Source&IAU name&Redshift&Balnicity Index&Outflow Velocity&Outflow $\sigma$\\
&&&(km~s$^{-1}$)&(km~s$^{-1}$)&(km~s$^{-1}$)\\
\hline\\
%&&&&&\\
\balqsoashort&\balqsoa&$2.419^{+0.004}_{-0.001}$&$640\pm160$&$-2320\pm70$&$560\pm60$\\%X21211_017
\balqsobshort&\balqsob&$2.479^{+0.006}_{-0.002}$&$120\pm190$&$-2800\pm120$&$880\pm80$\\%X21211_104
\balqsocshort&\balqsoc&$3.115\pm 0.001$       &$30\pm40$&$-2630\pm80$&$1470\pm40$\\%X00955_00149
\balqsodshort&\balqsod&$2.314\pm 0.002$       &$890\pm160$&$-4200\pm130$&$1710\pm90$\\%X01135_00026
\balqsoeshort&\balqsoe&$3.288\pm 0.001$       &$2450\pm110$&$-2090\pm30$&$700\pm30$\\%X21565_00025
\balqsofshort&\balqsof&$3.524\pm 0.001$       &$1590\pm110$&$-14500\pm100$&$1750\pm80$\\%X00912_00137
\end{tabular}
\end{center}
\end{table*} 

The sample is drawn from the \xmm\ Wide Angle Survey \citep[XWAS;
][]{esquej13}, an optical identification programme for X-ray sources,
carried out using the 2dF spectrograph \citep{lewis02} on the
Anglo-Australian Telescope. The first stage of sample selection
involved a visual inspection of XWAS QSOs to identify those which have
deep absorption lines blueward of C\,IV $\lambda$1550. Next,
absorption features and parts of the spectra shortward of N\,V
$\lambda$1250 were masked, before cross correlating with the QSO
template spectrum of \citet{vandenberk01} to obtain redshifts of the
objects. Note that for these strong-absorption objects we expect the
redshifts obtained by this process to be more accurate than those
derived in \citet{esquej13}. The sample was then refined by measuring
the BI of each candidate object. 
Following \citet{weymann91}, we consider any
object with BI~$>$0 to be a BALQSO, and only such objects
were included in the final sample, which consists of 6 objects.  Their
properties are summarised in Table~\ref{tab:opticalproperties}, and their 
optical spectra, as measured in XWAS, are shown in Fig.~\ref{fig:grest}.

\section{X-ray data and construction of X-ray spectra}

X-ray spectra were constructed for the BALQSOs using all the 
{\em XMM-Newton} European Photon Imaging Camera 
\citep[EPIC;][]{struder01, turner01} observations available in the 
{\em XMM-Newton} Science Archive. The observations are listed in
Table~\ref{tab:obsids}. The observations were reduced using standard
tasks in the {\em XMM-Newton} {\sc standard analysis software (sas)}
version 14.0\footnote{http://xmm.esac.esa.int/sas}, 
with periods of high background filtered out. Spectra
were accumulated from each observation using elliptical source regions
with the major axis sized according to the off-axis angle and oriented
perpendicular to the off-axis vector to match the point spread
function. Background was taken from an annular region around the
source, from which circular regions around other sources were
excised. Response and effective area files were then constructed for
each source and for each EPIC camera, in each observation, using the
standard {\sc sas} tasks {\sc rmfgen} and {\sc arfgen}. Then, the
source spectra, background spectra, responses and effective areas for
all EPIC cameras and all observations were combined to produce a
single source spectrum, background spectrum, response and effective
area file for each BALQSO, following the procedure described in
\citet{page03}. Spectra were grouped to a minimum of 20 counts per bin
using {\sc grppha}\footnote{http://heasarc.nasa.gov/docs/software/ftools}. 

\begin{table*}
\begin{center}
\caption{Galactic column densities for the BALQSOs and {\em XMM-Newton} observations used in the construction of the X-ray spectra. Column densites are taken from the Leiden/Argentine/Bonn Survey of Galactic H\,I \citep{kalberla05}. The column labelled ``Counts'' gives the total number of net (i.e. background-subtracted) 0.2--12~keV counts used in the combined EPIC spectrum for each source.}
\label{tab:obsids}
\begin{tabular}{cccccccccc}
Source&$N_{H}$&Counts&Obs ID&\multicolumn{3}{c}{Filters}&\multicolumn{3}{c}{Exposure time (ks)}\\
&(10$^{20}$cm$^{-2}$)&&&pn&MOS1&MOS2&pn&MOS1&MOS2\\
\hline\\
%&&&&&&&&\\
\balqsoashort&2.58&197&0123920101&Medium&-&-&14.9&-&-\\%X21211_017
        &&&0125320401&Medium&Medium&Medium&17.0&21.4&21.4\\%X21211_017
        &&&0125320501&Thin&Medium&Medium&3.5&6.3&6.2\\%X21211_017
        &&&0125320701&Medium&Thin&Thin&10.5&14.8&14.7\\%X21211_017
        &&&0133120301&Thin&Thin&Thin&6.5&10.7&10.7\\%X21211_017
        &&&0133120401&Thin&Thin&Thin&7.3&10.7&10.6\\%X21211_017
        &&&0153950101&-&Thin&Thin&-&4.5&4.5\\%X21211_017
&&&&&&&&&\\
\balqsobshort&2.39&133&0123920101&Medium&-&-&15.4&-&-\\%X21211_104
        &&&0125320401&Medium&Medium&Medium&17.0&21.4&21.4\\%X21211_104
        &&&0125320501&Thin&Medium&Medium&3.5&6.2&6.2\\%X21211_104
        &&&0125320701&Medium&Thin&Thin&10.5&14.8&14.7\\%X21211_104
        &&&0133120301&Thin&Thin&Thin&6.5&10.7&10.7\\%X21211_104
        &&&0133120401&Thin&Thin&Thin&7.3&10.7&10.6\\%X21211_104
        &&&0153950101&Medium&-&Thin&1.7&-&4.5\\%X21211_104
&&&&&&&&&\\
\balqsocshort&1.54&341&0112300101&Thin&Thin&Thin&16.1&24.6&25.7\\%X00955_00149
        &&&0602010101&Thin&Thin&Thin&71.1&87.3&87.7\\%X00955_00149
&&&&&&&&&\\
\balqsodshort&1.29&72&0109540101&Medium&Medium&Medium&42.0&53.0&52.9\\%X01135_00026
&&&&&&&&\\
\balqsoeshort&2.31&78&0112681301&Thin&Thin&Thin&9.6&16.7&16.7\\%X21565_00025
&&&&&&&&\\
\balqsofshort&2.61&237&0110980801&Thin&Thin&Thin&30.1&39.0&39.2\\%X00912_00137
\end{tabular}
\end{center}
\end{table*} 

\section{Results}
\subsection{X-ray spectral analysis}
\label{sec:xspecs}

\begin{figure*}
\begin{center}
\leavevmode
\hspace{-8mm}
\psfig{figure=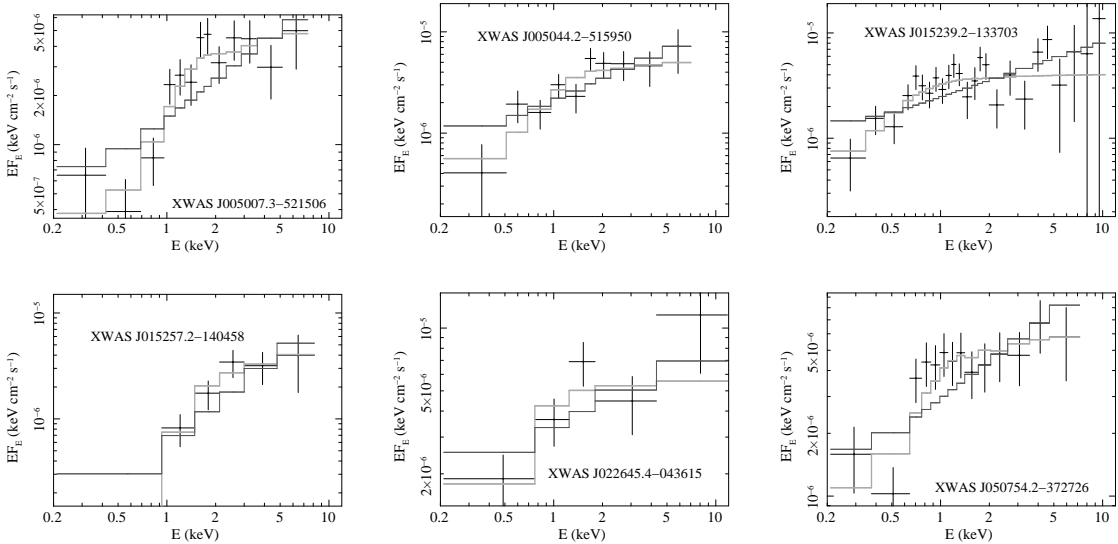,width=150mm}
\caption{{\em XMM-Newton} EPIC spectra of the six BALQSOs. The spectra
  have been flux calibrated by dividing through by the product of the
  effective area and Galactic transmission and are shown as
  $EF_{E}$. A power law with $\alpha = 1$ would be a horizontal line. The dark stepped lines show the power law model (model PL in Table~\ref{tab:fitting}) while the light-grey stepped lines show the model including an ionized absorber (model PL~$\times$~IA in Table~\ref{tab:fitting}).}
\label{fig:xspecs}
\end{center}
\end{figure*}

The X-ray spectra of the six BALQSOs are shown in
Fig.~\ref{fig:xspecs}.  Analyses of the X-ray spectra were carried out
in the spectral fitting package {\sc spex} version
2.02.04\footnote{http://www.sron.nl/divisions/hea/spex/index.html}. As
a first step the spectra were fitted using a power law model, with
energy index $\alpha$ free to vary, and photoelectric absorption from
a fixed column density of cold Galactic gas, determined from the
Leiden/Argentine/Bonn H\,I survey \citep{kalberla05}. The results are
given in Table \ref{tab:fitting}. The best fit slopes $\alpha$ range
from $-0.49$ to 0.50, significantly below the mean $\langle \alpha
\rangle = 0.98$ for X-ray selected QSOs \citep{mateos05}, indicating
that the BALQSOs have harder spectral shapes than are usual for
QSOs. For five of the six BALQSOs the simple power law fit provides an
acceptable goodness of fit (null hypothesis probability $>1$~per
cent), but the fit is poor for \balqsoashort.

A deficiency of soft X-rays is commonly a sign of absorption, so we
have considered an alternative model consisting of a power law with
ionized absorption as well as absorption from cold Galactic
material. As before, the Galactic $N_{H}$ was fixed. This time, we
fixed the power law slope to $\alpha=0.98$. For the ionized
absorption, we used the {\sc xabs} model in {\sc spex}, which predicts
both bound-bound and bound-free absorption from an equilibrium
photo-ionized plasma for which the ionization state is characterised
by the ionization parameter $\xi$. The {\sc xabs} model assumes 
an ionizing continuum shape based on the type 1 AGN 
NGC~5548 \citep{steenbrugge05} and a  
Gaussian velocity dispersion for the absorber. With respect to the 
rest-frame of the QSO, we assume that the
absorber seen in the X-ray band has a similar outflow velocity and velocity
dispersion as the UV absorber. Therefore, appropriate parameters were
determined from Gaussian fits to the main C~IV absorbing trough in
each restframe UV spectrum, and are given in
Table~\ref{tab:opticalproperties}.

\begin{table*}
\begin{center}
  \caption{Results of spectral fitting. The models are a power law
    (labelled PL), and a power law attenuated by an ionized absorber,
    with the power law index fixed at $\alpha_{X}=0.98$ (labelled PL
    $\times$ IA). In both cases a photoelectric absorber with fixed
    column density was included to represent Galactic absorption. 
$P$ is the null hypothesis probability corresponding to $\chi^{2}/\nu$. An asterix is used to indicate parameter uncertainties for which the limits of the fitting range were reached before $\Delta~\chi^{2}=1$.}
\label{tab:fitting}
\begin{tabular}{lcccccc}
&\balqsoashort&\balqsobshort&\balqsocshort&\balqsodshort&\balqsoeshort&\balqsofshort\\
%&XWAS~J005007&XWAS~J005044&XWAS~J015239&XWAS~J015257&XWAS~J022645&XWAS~J050754\\
\hline\\
PL&&&&&&\\
$\alpha_{X}$  &$0.12\pm0.10$&$0.14\pm0.14$&$0.47\pm0.08$&$-0.45\pm0.20$&$0.39\pm0.18$&$0.40\pm0.09$\\
$\chi^{2}/\nu$&26.4/11&9.6/8&37.9/23&8.9/4&4.4/3&22.5/12\\
$P$&0.0057&0.29&0.026&0.064&0.22&0.032\\
&&&&&&\\
PL $\times$ IA&&&&&&\\
log $\xi$&$2.8^{+0.3}_{-0.5}$&$2.6^{+0.4}_{-2.6*}$&$0.0^{+2.5}_{-0.0*}$&$2.6^{+0.6}_{-2.6*}$&$1.1^{+2.7}_{-1.1*}$&$2.8^{+0.4}_{-2.8*}$\\
log $N_{H}$&$23.6^{+0.3}_{-0.3}$&$23.3^{+0.4}_{-0.5}$&$22.5^{+0.5}_{-0.1}$&$23.9^{+0.4}_{-0.6}$&$22.7^{+0.6}_{-0.3}$&$23.4^{+0.4}_{-0.6}$\\
$\chi^{2}/\nu$&8.8/10&6.9/7&21.7/22&1.0/3&3.3/2&9.3/11\\
$P$&0.56&0.43&0.48&0.31&0.19&0.60\\
&&&&&&\\
\end{tabular}
\end{center}
\end{table*} 

\begin{figure*}
\begin{center}
\leavevmode
\hspace{-8mm}
\psfig{figure=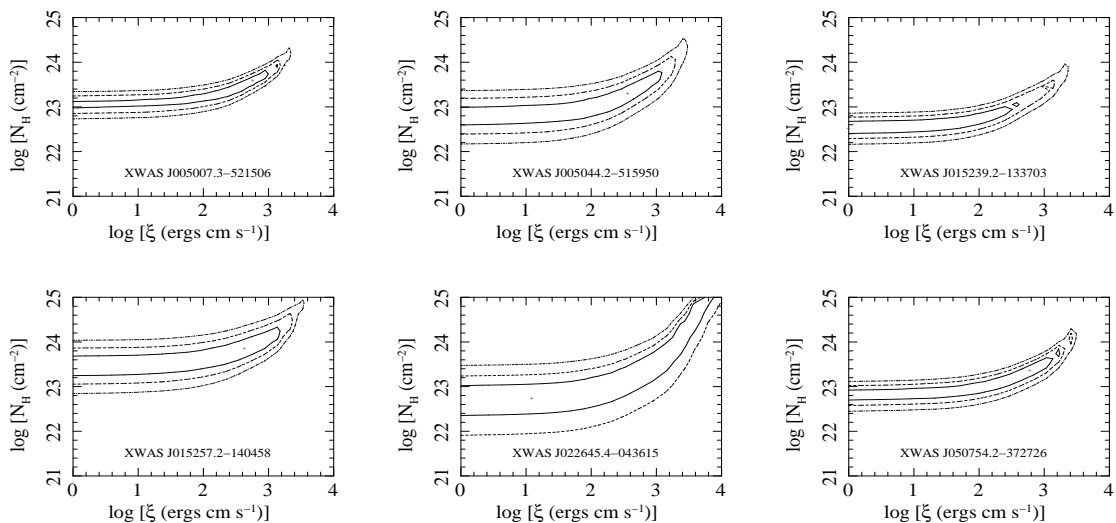,width=150mm}
\caption{Confidence contours for $\xi$ and $N_{H}$ in the ionized absorber model fits to the {\em XMM-Newton} spectra. The small cross indicates the best-fitting parameter values. Solid, dashed and dot-dashed contours correspond to 68, 95 and 99.73 per cent confidence for 2 interesting parameters.}
\label{fig:xconts}
\end{center}
\end{figure*}

The power law and ionized absorber model
provides acceptable fits for all 6 BALQSO spectra. Confidence contours
for log~$\xi$ and log~$N_{H}$ of the absorbers, as derived from the
X-ray fits are shown in Fig.~\ref{fig:xconts}. The ionization
parameters of the absorbers are poorly constrained from the fits, and
somewhat correlated with the column densities. Nonetheless, robust lower
limits on the column densities of between $10^{22}$ and
$10^{23}$~cm$^{-2}$ can be inferred from Fig.~\ref{fig:xconts} for all
of the objects except \balqsoeshort, which has the poorest quality
X-ray spectrum. 

\subsection{X-ray to optical ratio}
\label{sec:xrayopt}

Following the usual convention in AGN studies, we have parameterised
the X-ray to optical flux ratio of our BALQSOs as $\alpha_{OX}$, the
power law slope which would connect the flux densities at 2500~\AA\
and 2~keV in the rest-frame of the source. The 2~keV flux measurements
were derived from the power law model fits to the X-ray spectra
without ionised absorbers (i.e. model PL in Table~\ref{tab:fitting}),
similar to the approach taken by \citet{gallagher06} and
\citet{blustin08}. The rest-frame 2500~\AA\ flux was derived from $R$
band magnitudes measured by the SuperCOSMOS Sky Survey
\citep{hambly01a} scans of the UK Schmidt Telescope $R$ plates,
converted to monochromatic flux densities assuming the template
spectrum of \citet{vandenberk01}. Uncertainties on
the R magnitudes are taken from Table~6 of \citet{hambly01b}. 
Uncertainties on $\alpha_{OX}$
were derived by propagating the statistical uncertainties on the X-ray
model fits and R magnitudes.
The values of $\alpha_{OX}$ are given in Table~\ref{tab:alphaox}, together with 
broadband fluxes and magnitudes. 

\begin{table*}
\begin{center}
  \caption{Magnitudes, broadband fluxes and $\alpha_{OX}$. The $R$
    magnitudes come from the SuperCOSMOS Sky Survey scans of the UK
    Schmidt Telescope $R$ plates. Extinction estimates come from the
    \citet{schlafly11} calibration of the \citet{schlegel98} dust
    maps. The 0.5-2.0 keV flux values are derived from the power law
    spectral fits without any intrinsic absorption 
    (model PL in Table~\ref{tab:fitting}). The column
    headed `$\alpha_{OX}$ observed' gives the values of $\alpha_{OX}$
    in which the rest-frame 2~keV flux is derived from the power law
    fits (model PL in Table~\ref{tab:fitting}), while the column
    headed `$\alpha_{OX}$ corrected' gives values of $\alpha_{OX}$ in
    which the rest-frame 2~keV flux is derived from the power law
    continuum in the ionized absorber model (model PL $\times$ IA in
    Table~\ref{tab:fitting}), with no attenuation from the ionized
    absorption. The column headed `$\alpha_{OX}$ predicted' gives the
    values of $\alpha_{OX}$ that would be predicted from the
    2500~\AA\ monochromatic luminosities and equation 3 from
    \citet{just07}.}
\label{tab:alphaox}
\begin{tabular}{ccccccc}
Source&$R$&$A_{R}$&0.5-2.0 keV flux &$\alpha_{OX}$&$\alpha_{OX}$&$\alpha_{OX}$\\
&(mag)&(mag)&(10$^{-15}$~erg~cm$^{-2}$~s$^{-1}$)&observed&corrected&predicted\\
\hline\\
%&&&&\\
\balqsoashort&$19.51\pm 0.13$&0.04&$3.25\pm 0.34$&$1.75\pm 0.03$&$1.41^{+0.08}_{-0.07}$&$1.62$\\
\balqsobshort&$20.69\pm 0.18$&0.04&$4.70\pm 0.62$&$1.49\pm 0.04$&$1.20^{+0.08}_{-0.16}$&$1.56$\\
\balqsocshort&$19.14\pm 0.13$&0.04&$5.57\pm 0.41$&$1.68\pm 0.03$&$1.52^{+0.03}_{-0.03}$&$1.67$\\
\balqsodshort&$20.44\pm 0.18$&0.03&$1.26\pm 0.32$&$1.87\pm 0.06$&$1.33^{+0.11}_{-0.25}$&$1.57$\\
\balqsoeshort&$18.87\pm 0.11$&0.06&$7.46\pm 1.10$&$1.67\pm 0.04$&$1.52^{+0.05}_{-0.13}$&$1.70$\\
\balqsofshort&$19.13\pm 0.13$&0.07&$6.68\pm 0.54$&$1.70\pm 0.03$&$1.46^{+0.06}_{-0.07}$&$1.69$\\
\end{tabular}
\end{center}
\end{table*} 

\section{Discussion}
\label{sec:discussion}

The X-ray spectra of the BALQSOs show significant evidence for
absorption. Although the majority of the X-ray spectra can be
successfully fitted with power-law models without intrinsic
absorption, the derived spectral indices ($-0.5 < \alpha_{X} < 0.5$)
are significantly lower than the norm for non-BAL QSOs, which cluster
around a mean of $\langle \alpha_{X} \rangle = 0.98$ with a dispersion
of $\sigma_{\alpha}=0.27$ \citep{mateos05,mateos10,scott11},
indicating that the BALQSOs have significantly harder shapes than
non-BALQSOs. All of the spectra can be fitted successfully with a
model in which a typical 
non-BAL
QSO spectrum passes through an
ionized absorber. The six sources presented here therefore have X-ray
spectral properties which are consistent with those of the five X-ray
selected BALQSOs studied by \citet{blustin08}.

On the other hand, the observed ultraviolet to X-ray spectral slopes
$\alpha_{OX}$ of our six X-ray selected BALQSOs are quite
consistent with those expected for non-BAL QSOs. There is a well-known
relation between ultraviolet luminosity and $\alpha_{OX}$ in non-BAL
QSOs \citep{strateva05,steffen06,just07}.  The relation holds for
  both optically-selected and X-ray selected non-BAL QSO
  samples \citep{lusso10}.  Using the expression from
\citet{just07}, $\alpha_{OX} = 0.14 (\log L_{\nu}) + 2.705$ where
$L_{\nu}$ is the monochromatic luminosity at 2500\AA, we have
calculated the predicted $\alpha_{OX}$ values (see Table
\ref{tab:alphaox}). Taking into account the observed scatter around
this relation of $\sigma_{\alpha} = 0.14$ \citep{just07}, we would
predict a mean of $\langle \alpha_{OX} \rangle = 1.64\pm 0.06$ for our
sample of objects
if they were non-BAL QSOs,
which is quite consistent with the observed
$\langle \alpha_{OX} \rangle = 1.69\pm 0.05$. In contrast,
optically-selected samples of BALQSOs usually show much larger values
of $\alpha_{OX}$ than our X-ray-selected sample. For example the
majority of the sample of 36 BALQSOs from the Large Bright Quasar
Survey studied by \citet{gallagher06} have $\alpha_{OX} > 2$, and
their $\langle \alpha_{OX} \rangle$ is larger by 0.52 than the mean
that would be expected given their ultraviolet luminosities.

At first sight, finding that our BALQSOs have $\alpha_{OX}$ values
in the normal range for 
non-BAL
QSOs seems at odds with the presence of
significant X-ray absorption in their X-ray spectra, because the X-ray
absorption would be expected to lead to increased $\alpha_{OX}$
compared to a typical QSO without significant X-ray absorption. An
explanation can be gleaned by examining how the absorption should
affect $\alpha_{OX}$. For each of our BALQSOs we have recalculated the
$\alpha_{OX}$ using the intrinsic (i.e. unabsorbed) rest-frame 2~keV
model continuum from the ionized absorber model fits in
Table~\ref{tab:fitting}. These absorption-corrected values are given
in Table~\ref{tab:alphaox} in the column headed ``$\alpha_{OX}$
corrected''. In all cases the absorption-corrected $\alpha_{OX}$
values are lower than those predicted by the \citet{just07} relation
by between 1 and 3 times the scatter (0.14) found by
\citet{just07}. Therefore intrinsically, these objects occupy the
lowest 15 per cent of the $\alpha_{OX}$ distribution. 

In constructing an X-ray selected BALQSO sample we have naturally
selected BALQSOs which are at the X-ray-bright end of the
distribution. As such we would expect to find the part of the
population with the lowest levels of X-ray absorption, and/or the
smallest intrinsic values of $\alpha_{OX}$. We do not find any
examples of BALQSOs without strong X-ray absorption, but all of our
sample have intrinsic $\alpha_{OX}$ values which are low compared to
the majority of QSOs at similar UV luminosities. 

Consistent with our findings, \citet{blustin08} also found evidence
for strong X-ray absorption in the X-ray spectra of all of their X-ray
selected BALQSOs. The values of $\alpha_{OX}$ found by
\citet{blustin08} are a little higher than for the six objects in our
sample, though still much lower than typically found for optically
selected BALQSOs. Observationally, the sample of \citet{blustin08} was
selected from deep X-ray surveys, whereas ours is drawn
serendipitously from moderate exposure ($<50$~ks) XMM-Newton
observations. The mean 0.5--2~keV flux of the \citet{blustin08} sample
is 5 times lower than the mean 0.5--2~keV flux of the six sources
studied here, and the mean 0.5--2~keV luminosity of the
\citet{blustin08} sample is 7 times lower than the mean 0.5--2~keV
luminosity of the six sources studied here.  By selecting BALQSOs at a
higher X-ray flux limit than \citet{blustin08}, we have been able to
push further into the X-ray-bright end of the BALQSO population, and
so are able to draw stronger conclusions regarding their X-ray
properties. Our findings suggest that X-ray absorption is a universal
property of BALQSOs, but imply that X-ray weakness (i.e. large
$\alpha_{OX}$) is not.

In the study of \citet{streblyanska10} there were two notable
correlations concerning the X-ray absorbers, which
\citet{streblyanska10} modelled as ionized absorbers. First, they
found a correlation between the ionization parameter $\xi$ and
$N_{H}$, and second they found a correlation between $N_{H}$ and the
BI of the ultraviolet absorption. Our sample is too small to permit
such a correlation analysis, but our spectral fitting may provide some
insights into these correlations. It is clear from
Fig.~\ref{fig:xconts} that with the current statistics in our X-ray
spectra, the ionization parameters of the absorbers are not
constrained, and there is significant covariance between log~$\xi$ and
log~$N_{H}$ in the individual spectral fits. The six BALQSOs in our 
sample show a consistent trend in correlation between log~$\xi$ and 
log~$N_{H}$ to that found by \citet{streblyanska10}, but it can be 
accounted for entirely by this covariance. 
The ionized absorption model we have used, {\sc xabs} in {\sc spex},
is more sophisticated than that used by \citet{streblyanska10}, {\sc
  absori} in {\sc xspec} \citep{arnaud96}, 
because {\sc xabs} includes bound-bound
transitions as well as bound-free transitions, while {\sc absori}
includes only bound-free transitions. However, we would expect a
similar level of covariance between log~$\xi$ and log~$N_{H}$ in the
two models.

Most of the BALQSOs in
the sample of \citet{streblyanska10} for which spectral fitting has
been performed are brighter in X-rays than those in our sample, so the
degeneracy between $\xi$ and $N_{H}$ may be mitigated to some extent
by the improved photon statistics compared to our sample, but it is
nonetheless a concern that some or all of the correlation found by
\citet{streblyanska10} between $\xi$ and $N_{H}$ may be a consequence
of the covariance of these two parameters. 

Of the correlation found by \citet{streblyanska10} between BI and
$N_{H}$ we can neither confirm nor refute this correlation with our small 
sample. 
Some insight on the relation between ultraviolet C\,IV absorption and
X-ray absorption in QSOs may be drawn by comparing our sample of
BALQSOs with the X-ray absorbed QSOs studied by
\citet{page11}. Comparison of Fig.~\ref{fig:xconts} with the
corresponding figure~4 in \citet{page11} (and disregarding the object
RX\,J124913 from their sample, which is itself a BALQSO), we find that
the X-ray absorbed QSOs have confidence contours in (log~$\xi$,
log~$N_{H}$) which fall systematically below those of our BALQSO
sample. Despite the degeneracy between $\xi$ and $N_{H}$ in both
samples, the X-ray selected BALQSOs have systematically more X-ray
absorption than the X-ray absorbed QSOs, which were also X-ray
selected. The X-ray absorbed QSOs 
of \citet{page11}
also have lower C\,IV equivalent
widths than the BALQSOs.  There thus appears to be a continous
spectrum of QSO absorption properties, from the majority of QSOs which
have C\,IV equivalent widths of $< 5$~\AA\ and minimal X-ray
absorption, through the X-ray absorbed QSOs which typically have C\,IV
equivalent widths in the range of $5-20$~\AA\ and moderate X-ray
absorption, to the BALQSOs which have C\,IV equivalent widths
$>20$~\AA\ and the strongest X-ray absorption. A similar correlation
between C\,IV equivalent width and $\alpha_{OX}$ has long been
established \citep{brandt00}. As X-ray absorption appears to be a
universal property of BALQSOs but large $\alpha_{OX}$ is not, the
connection between C\,IV absorption and X-ray absorption may be
considered more fundamental than that between C\,IV and $\alpha_{OX}$.

Significant X-ray absorption, with log~$N_{H}>10^{22.5}$~cm$^{-2}$
appears to be a ubiquitous property amongst X-ray selected BALQSOs,
defined according to the classical BI $> 0$ criterion, as presented
here and by \citet{blustin08}. The same conclusion was reached by
\citet{streblyanska10} for BALQSOs identified by correlating optical
catalogues with the 2XMM catalogue. 
We are thus led to the same conclusion as
\citet{streblyanska10} that the rather different absorption properties found by
\citet{giustini08} for X-ray detected BALQSOs, probably relates to
their use of the AI definition of BALQSOs. 
The AI criterion used by \citet{giustini08} includes objects with
lower C\,IV equivalent widths than the classical BI criterion, and
this may account for the comparatively weak X-ray absorption found by
those authors for some of their sample. The apparent difference in
X-rays between BALQSOs defined by BI and those defined by AI (and not
BI) is interesting in the context of the bimodality of the AI
distribution demonstrated by \citet{knigge08} and their interpretation
that the bimodality corresponds to two distinct sub-populations of
QSOs. It appears that the two sub-populations selected by AI have
distinct distributions in X-ray absorption, as well as in their UV
absorption-line properties.

\section{Conclusions}
\label{sec:conclusions}

We have presented a sample of six X-ray selected BALQSOs from
XWAS. Examination of their X-ray spectra indicates that all have hard
spectral shapes compared to the majority of QSOs, suggesting
absorption. An ionized absorption model applied to a typical AGN
template X-ray spectrum is able to reproduce the X-ray spectral
shapes, and implies that these are intrinsically rather X-ray bright
for QSOs given their UV luminosities.  X-ray absorption appears to be
ubiquitous in classically-defined (i.e. BI~$>$~0) BALQSOs, and this is
true for X-ray-selected BALQSOs as well as for optically-selected
samples. In contrast, X-ray weakness (i.e. high $\alpha_{OX}$) is not
a universal property of BALQSOs.  We have argued that there is a
continuous spectrum of X-ray absorption properties in QSOs which are
related to the equivalent width of ultraviolet C\,IV absorption, with
BALQSOs forming the most heavily absorbed end of the distribution.

\section{Acknowledgments}

Based on observations obtained with \xmm, an ESA science mission with
instruments and contributions directly funded by ESA Member States and
NASA.  This research was also based on observations made at the
Anglo-Astralian Telescope. MJP acknowledges financial support from the
UK Science and Technology Facilities Council. FJC and SM acknowledge
financial support through grant AYA2015-64346-C2-1-P (MINECO/FEDER).
MTC acknowledges support by the Spanish Programma Nacional de
Astronomia y Astrofisica under grant AYA2009-08059.  MK acknowledges
support by DFG grant KR 3338/3-1.  The NASA/IPAC Extragalactic
Database (NED) is operated by the Jet Propulsion Laboratory,
California Institute of Technology, under contract with the National
Aeronautics and Space Administration.


\begin{thebibliography}{}

\bibitem[Arnaud(1996)]{arnaud96}
Arnaud K.A., 1996,
Astronomical Data Analysis Software and Systems V, eds. 
Jacoby G. and Barnes J., ASP Conf. Series volume 101, p17

\bibitem[Barcons et~al.(2002)]{barcons02}
Barcons X., et~al., 2002,
A\&A, 382, 522

\bibitem[Blustin et~al.(2008)]{blustin08}
Blustin A.J., et~al., 2008, MNRAS, 390, 1229

\bibitem[Brand, Laor \& Wills(2000)]{brandt00}
Brandt W.N., Laor A. \& Wills B.J., 
2000, ApJ, 528, 637

\bibitem[Esquej et~al.(2013)]{esquej13}
Esquej P., et~al., 
2013, A\&A, 577, A123

\bibitem[Gallagher et~al.(2006)]{gallagher06}
Gallagher S.C., Brandt W.N., Chartas G., Priddey R., Garmire G.P., 
Sambruna R.M., 
2006, ApJ, 644, 709

\bibitem[Giacconi et~al.(2001)]{giacconi01}
Giacconi R., et~al.,
2001, ApJ, 551, 624

\bibitem[Gibson et~al.(2009)]{gibson09}
Gibson R.R., et~al., 
2009, ApJ, 692, 758

\bibitem[Giustini et~al.(2008)]{giustini08}
Giustini M., Cappi M., Vignati C.,
2008, A\&A, 491, 425

\bibitem[Green et al.(1995)]{green95}
Green P.J., 1995, ApJ, 450, 51

\bibitem[Green \& Mathur(1996)]{green96}
Green P.J. \& Mathur S., 1996, ApJ, 462, 637

\bibitem[Hall et~al.(2002)]{hall02}
Hall P.B., et~al.,
2002, ApJS, 141, 267

\bibitem[Hambly et~al.(2001a)]{hambly01a}
Hambly N.C., et~al.,
2001, MNRAS, 326, 1279

\bibitem[Hambly et~al.(2001b)]{hambly01b}
Hambly N.C., Irwin M.J., MacGillivray H.T., 
2001, MNRAS, 326, 1295

\bibitem[Just et~al.(2007)]{just07}
Just D.W., Brandt W.N., Shemmer O., Steffen A.T., Schneider D.P., 
Chartas G., Garmire G.P., 
2007, ApJ, 665, 1004

\bibitem[Kalberla et~al.(2005)]{kalberla05}
Kalberla P.M.W., Burton W.B., Hartman Dap, Arnal E.M., 
Bajaja E., Morras R., P\"oppel W.G.L., 
2005, A\&A, 440, 775

\bibitem[Knigge et~al.(2008)]{knigge08}
Knigge C., Scaringi S., Goad M.R., Cottis C.E.,
2008, MNRAS, 386, 1426 

\bibitem[Lewis et~al.(2002)]{lewis02}
Lewis I.J., et~al., 
2002, MNRAS, 333, 279

\bibitem[Lusso et~al.(2010)]{lusso10}
Lusso E., et~al., 2010, A\&A, 512, A34

\bibitem[Mateos et~al.(2005)]{mateos05}
Mateos S., et~al., 2005, A\&A, 433, 855

\bibitem[Mateos et~al.(2010)]{mateos10}
Mateos S., et~al., 2010, A\&A, 510, 35

\bibitem[Morabito et~al.(2014)]{morabito14}
Morabito L.K., Dai X., Leighly K.M., Sivakoff G.R., Shankar F.,
2014, ApJ, 786, 58

\bibitem[Page, Davis \& Salvi(2003)]{page03}
Page M.J., Davis S.W. \& Salvi N.J., 2003, MNRAS, 343, 1241

\bibitem[Page et~al.(2011)]{page11}
Page M.J., Carrera F.J., Stevens J.A., Ebrero J., Blustin A.J., 
2011, MNRAS, 416, 2792

\bibitem[Scaringi et~al.(2009)]{scaringi09}
Scaringi S., Cottis C.E., Knigge C., Goad M.R.,
2009, MNRAS, 399, 2231

\bibitem[Schlafly \& Finkbeiner(2011)]{schlafly11}
Schlafly E.F. \& Finkbeiner D.P., 2011, ApJ, 737, 103

\bibitem[Schlegel et al.(1998)]{schlegel98}
Schlegel D.J., Finkbeiner D.P., Davis M., 1998, ApJ, 500, 525

\bibitem[Schneider et~al.(2007)]{schneider07}
Schneider D.P., et~al., 
2007, AJ, 134, 102

\bibitem[Scott et~al.(2011)]{scott11}
Scott A., Stewart G.C., Mateos S., Alexander D.M., Hutton S., Ward M.J.,
2011, MNRAS, 417, 992

\bibitem[Steenbrugge et~al. (2005)]{steenbrugge05}
Steenbrugge K.C., et~al., 
2005, A\&A, 434, 569

\bibitem[Steffen et~al.(2006)]{steffen06}
Steffen A.T., Strateva I., Brandt W.N., Alexander D.M., KoekemoerA.M., 
Lehmer B.D., Schneider D.P., Vignali C., 
2006, AJ, 131, 2826

\bibitem[Strateva et~al.(2005)]{strateva05}
Strateva I.V., Brandt W.N., Schneider D.P., Vanden Berk D.G., Vignali C.,
2005, AJ, 130, 387 

\bibitem[Streblyanska et~al.(2010)]{streblyanska10}
Streblyanska A., Barcons X., Carrera F.J., Gil-Merino R.,
2010, A\&A, 515, A2

\bibitem[Str\"uder et~al.(2001)]{struder01}
Str\"uder L., et~al., 2001, A\&A, 365, L18 

\bibitem[Trump et~al.(2006)]{trump06}
Trump J.R., et~al.,
2006, ApJS, 165, 1

\bibitem[Turner et~al.(2001)]{turner01}
Turner M.J., et~al., 2001, A\&A, 365, L27

\bibitem[Vanden Berk et~al.(2001)]{vandenberk01}
Vanden Berk D.E., et~al., 2001, AJ, 122, 549

\bibitem[Watson et~al.(2009)]{watson09}
Watson M.G., et~al., 
2009, A\&A, 493, 339

\bibitem[Weymann et~al.(1991)]{weymann91}
Weymann R.J., Morris S.L., Foltz C.B., Hewett, P.C.,
1991, ApJ, 373, 23


\end{thebibliography}
\end{document}